# Collective Iterative Learning Control: Exploiting Diversity in Multi-Agent Systems for Reference Tracking Tasks

Michael Meindl, Fabio Molinari, Dustin Lehmann, and Thomas Seel

*Abstract*—Multi-agent systems (MASs) can autonomously learn to solve previously unknown tasks by means of each agent's individual intelligence as well as by collaborating and exploiting collective intelligence. This article considers a group of autonomous agents learning to track the same given reference trajectory in a possibly small number of trials. We propose a novel collective learning control method that combines iterative learning control (ILC) with a collective update strategy. We derive conditions for desirable convergence properties of such systems. We show that the proposed method allows the collective to combine the advantages of the agents' individual learning strategies and thereby overcomes trade-offs and limitations of single-agent ILC. This benefit is achieved by designing a heterogeneous collective, i.e., a different learning law is assigned to each agent. All theoretical results are confirmed in simulations and experiments with two-wheeled-inverted-pendulum robots (TWIPRs) that jointly learn to perform the desired maneuver.

## I. Introduction

INTELLIGENT multi-agent systems (MASs) describe groups of interacting autonomous agents that have gathered interest in the scientific community and have also become an integral concept in industrial technology and robotics [2], [3]. Examples of MAS include schools of fish [4], autonomous cars at an intersection [5], or a swarm of mobile robots in a warehouse management system [6]. Such systems can autonomously learn to solve previously unknown tasks by means of each agent's individual intelligence as well as by collaborating and exploiting collective intelligence.

The tasks that intelligent MAS learn to solve can be distinguished into two types, depending on the characteristics of the required interaction. On the one hand, there are combined-action tasks, in which a collective goal can only be achieved by the combined actions of multiple agents. Examples include consensus problems, area coverage tasks, and lifting objects that are too heavy for a single agent. On the other hand, there are independent-action tasks, which can be decomposed into a number of identical or similar subtasks that can be solved by agents independently. Examples include a team of humans building a brick wall, a flock of drones exploring unknown territory or maneuvering through a narrow corridor, and a group of quadruped robots crossing an icy surface. In many real-world applications, MASs are faced with problems that contain both types of tasks, such as a team of manufacturing robots that independently assemble components of a product and then assemble the whole.

While combined-action tasks strictly require collaboration during the learning process, independent-action tasks can be addressed by letting all agents learn their subtask independently. The simplicity of this approach seems seductive, but it raises the question whether that is a reasonable strategy or whether intelligent MAS can benefit from collaboration and knowledge exchange also in independent-action tasks. If yes, then which form of cooperation is advantageous? Can the collective obtain emergent properties that none of the individuals can achieve on its own?

The present manuscript addresses these questions in the context of independent learning control tasks, in which each agent must learn to perform the same predefined motion or maneuver, as illustrated in Fig. 1. We extend the concept of iterative learning control (ILC) to the multi-agent case and propose a collective ILC (CILC) scheme that exploits the principles of diversity and cooperation. It is demonstrated by simulations and experiments that the proposed CILC approach merges different advantages of individual learning strategies and thereby overcomes performance trade-offs and limitations that are insuperable in single-agent ILC.

### A. Related Work and Contributions

In recent years, a number of different learning problems in MASs were studied and solved using learning control

Manuscript received June 3, 2021; accepted August 1, 2021. Manuscript received in final form August 31, 2021. This work was supported in part by the Deutsche Forschungsgemeinschaft (DFG), German Research Foundation, through Germany's Excellence Strategy-EXC 2002/1 "Science of Intelligence," under Project 390523135, in part by the German Research Foundation (DFG) through the Priority Programme SPP1914 "Cyber-Physical Networking" under Grant RA516/12-1, and in part by the Verbund der Stifter through the Project "Robotic Zoo." Recommended by Associate Editor Y. Pan. *(Corresponding author: Michael Meindl.)*

Michael Meindl is with the Embedded Mechatronics Laboratory, Hochschule Karlsruhe, 76133 Karlsruhe, Germany (e-mail: michael.meindl@hs-karlsruhe.de).

Fabio Molinari and Dustin Lehmann are with the Control Systems Group, Technische Universität Berlin, 10587 Berlin, Germany (e-mail: molinari@control.tu-berlin.de; lehmann@control.tu-berlin.de).

Thomas Seel is with the Department of Artificial Intelligence in Biomedical Engineering, Friedrich-Alexander-Universität Erlangen–Nürnberg, 91052 Erlangen, Germany (e-mail: thomas.seel@fau.de).

Color versions of one or more figures in this article are available at https://doi.org/10.1109/TCST.2021.3109646.

Digital Object Identifier 10.1109/TCST.2021.3109646

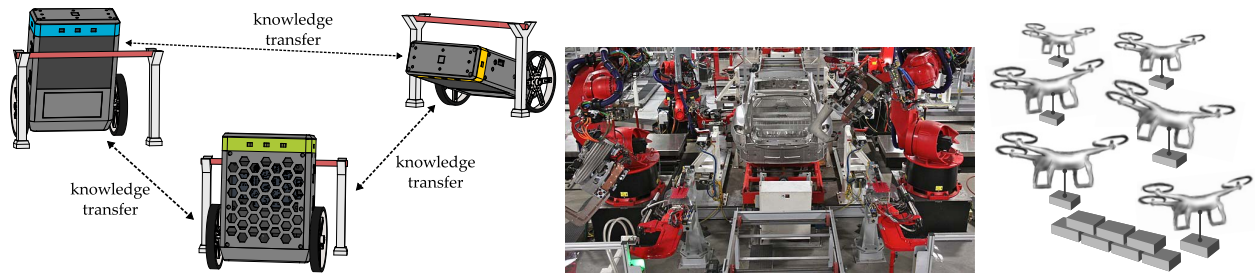

Fig. 1. Examples of independent-action tasks. (Left) A group of two-wheeled inverted pendulum robots (TWIPRs) has to dive beneath an obstacle. (Center) Multiple robots have to solve an identical manufacturing task in parallel (photograph by Steve Jurvetson licensed under CC BY 2.0). (Right) In a swarm of quadrotors, each drone has to deliver a brick to build a wall (see [1]).

methods as well as methods based on machine learning [7]. Most literature has focused on combined-action tasks, such as formation control, synchronization, or flocking in MAS (see [8]–[13]). In [14], a group of agents employs a distributed ILC method to follow a desired trajectory while holding a specified formation. A similar approach was taken in [15], where each agent uses ILC to iteratively reduce deviations from the desired formation along a given reference trajectory. Consensus-based ILC for distributed leader–follower problems has been investigated in [16] and [17]. ILC has also been used for more general consensus problems, in which the desired trajectory is only known to a subset of agents [18]. A more thorough literature review of learning control approaches for synchronization and formation tasks can be found in [19].

The potential benefits of cooperation are less obvious in independent-action tasks. To the best of our knowledge, there exists only one previous contribution that investigates the use of ILC methods for solving a multi-agent task that can be decomposed into independent subtasks: In [20], $N$ agents simultaneously solve the same disturbance estimation task while sharing their local information with the group. If all $N$ agents share their measurements with all others, then everybody has $N$ times more data to use for the estimation, which then converges (in the best case) $N$ times faster. A conceptually similar approach has been considered for the identification of nonlinear system dynamics by neural networks [21].

Despite these first results, many questions remain unanswered. Most remarkably, the role of heterogeneity and its potentially advantageous effect on the collective learning dynamics in independent-action tasks remains unclear. Across and beyond the aforementioned types of tasks, the vast majority of prior contributions in MAS considered groups of agents with similar or equal learning strategies and did not explore the possibility of heterogeneously designed agents. Finally, there are to date no simulative or even experimental investigations of the potential benefit of cooperation and the resulting distributed learning dynamics.

The present contribution proposes a distributed learning approach, called CILC, that enables MASs to solve independent-action tasks in which all agents simultaneously learn to perform the same desired maneuver. The learning law of each agent is designed differently, and each agent shares its measurements and control inputs with the collective during the learning, while likewise exploiting the received information. We derive conditions for monotonic convergence above a threshold and asymptotic stability in CILC systems, and we validate the method by simulations and by experiments using TWIPRs that learn to perform the desired maneuver. Heterogeneity in the agents' learning laws is found to be advantageous in the sense that the proposed method combines the benefits of individual learning laws and thereby overcomes performance tradeoffs and limitations that are insuperable in single-agent ILC.

The remainder of this article is structured as follows. Section II provides a brief review of some basic ILC concepts and results. The multi-agent learning control problem is formulated in Section III, and the novel approach to this problem is introduced in Section IV. Section V is dedicated to validating theoretical results and investigating the collective learning dynamics by simulations and real-world experiments. Conclusions and directions for future work are then given in Section VI.

*B. Notation*

Let $\mathbb{N}_0$ and $\mathbb{N}$ denote the set of nonnegative, respectively, positive, integers. Let $\mathbb{R}$ denote the set of real numbers and $\mathbb{R}_{>0}$, respectively, $\mathbb{R}_{\geq 0}$, the set of positive, respectively, nonnegative, real numbers. Vectors are in bold type and lower-case letters, e.g., **v**. Matrices are in bold type and upper-case letters, e.g., **A**. Indices of vectors are used to denote the ILC trial, e.g., $\mathbf{v}_j$ denotes the vector **v** on the $j$th ILC trial. Let $\|\mathbf{v}\|$ denote a norm of the vector **v**, and $\|\mathbf{A}\|$ the corresponding induced matrix norm of the matrix **A**. A particular example is the Euclidean norm denoted by $\|\cdot\|_2$. The spectral radius of a square matrix **A** is denoted by $\rho(\mathbf{A})$.

## II. ILC IN A NUTSHELL

ILC is a learning control strategy that solves reference tracking tasks by updating a feed-forward control input based on the measurements and the control inputs of previous trials [22]–[24]. Consider a repetitive, single-input, single-output, discrete-time system with linear dynamics, and a trial duration of $N \in \mathbb{N}$ sampling intervals. The system's input, respectively, output, variable at trial $j \in \mathbb{N}_0$ and sampling interval $n \in [1, N]$ is denoted by $u_j(n)$, respectively, $y_j(n)$. On each trial, the samples of the input and output variables are collected in the so-called input trajectory $\mathbf{u}_j \in \mathbb{R}^N$, respectively, output trajectory $\mathbf{y}_j \in \mathbb{R}^N$, with

$$\mathbf{u}_j := [u_j(1) \quad u_j(2) \quad \ldots \quad u_j(N)]^T \quad (1)$$
$$\mathbf{y}_j := [y_j(1+m) \quad y_j(2+m) \quad \ldots \quad y_j(N+m)]^T \quad (2)$$

where $m \in \mathbb{N}_0$ is the system's relative degree. The lifted form of the plant dynamics is given by

$$\forall j \in \mathbb{N}_0, \quad \mathbf{y}_j = \mathbf{P}\mathbf{u}_j + \mathbf{d} \quad (3)$$

where $\mathbf{P} \in \mathbb{R}^{N \times N}$ is the regular plant matrix and $\mathbf{d} \in \mathbb{R}^N$ is an unknown yet trial-invariant disturbance (see [25]).

The control task consists in letting the output $\mathbf{y}_j$ follow a desired reference trajectory $\mathbf{r} \in \mathbb{R}^N$. The tracking error is determined after each trial, that is,

$$\forall j \in \mathbb{N}_0, \quad \mathbf{e}_j := \mathbf{r} - \mathbf{y}_j. \quad (4)$$

The learning task is to update the input trajectory $\mathbf{u}_j$ from trial to trial such that a suitable norm of $\mathbf{e}_j$ decays rapidly. Here, we consider the commonly used learning law

$$\forall j \in \mathbb{N}_0, \quad \mathbf{u}_{j+1} = \mathbf{Q}(\mathbf{u}_j + \mathbf{L}\mathbf{e}_j) \quad (5)$$

where $\mathbf{L} \in \mathbb{R}^{N \times N}$, $\mathbf{Q} \in \mathbb{R}^{N \times N}$ denote the learning gain matrix and the Q-filter matrix [25].

*Definition 1 (Asymptotic Stability, see [25]):* A system with dynamics (3) and update law (5) is asymptotically stable if the limit

$$\lim_{j \to \infty} \mathbf{e}_j := \mathbf{e}_R \quad (6)$$

exists and is unique for any choice of $\mathbf{u}_0$.

*Theorem 1:* A system with dynamics (3) and update law (5) is asymptotically stable if

$$\rho(\mathbf{Q}(\mathbf{I} - \mathbf{L}\mathbf{P})) < 1. \quad (7)$$

*Proof:* See [25]. ∎

Asymptotic stability guarantees that the error trajectory converges to a final trajectory $\mathbf{e}_R$. However, the error may still grow during the learning process.

*Definition 2 (Monotonic Convergence, see [25]):* A system with dynamics (3) and update law (5) is monotonically convergent under a norm $\|\cdot\|$ with the so-called convergence rate $\gamma \in \mathbb{R}_{\geq 0}$ if $\gamma < 1$ and

$$\forall j \in \mathbb{N}_0, \quad \|\mathbf{e}_{j+1} - \mathbf{e}_R\| \leq \gamma \|\mathbf{e}_j - \mathbf{e}_R\|. \quad (8)$$

*Theorem 2:* A system with dynamics (3) and update law (5) is monotonically convergent under a norm $\|\cdot\|$ if

$$\gamma := \|\mathbf{P}\mathbf{Q}(\mathbf{I} - \mathbf{L}\mathbf{P})\mathbf{P}^{-1}\| < 1. \quad (9)$$

*Proof:* See [25]. ∎

*Remark 1:* Since the spectral radius of a square matrix is upper-bounded by any norm of that matrix, monotonic convergence implies asymptotic stability.

While (8) does not state whether the tracking error norm $\|\mathbf{e}_j\|$ decreases from trial to trial, the following does.

*Definition 3 (Monotonic Convergence Above a Threshold):* A system with dynamics (3) and update law (5) is monotonically convergent above a threshold $\kappa \in \mathbb{R}_{>0}$ under a given norm $\|\cdot\|$ if

$$\forall j \in \mathbb{N}_0, \quad \|\mathbf{e}_j\| \geq \kappa \implies \|\mathbf{e}_{j+1}\| \leq \|\mathbf{e}_j\|. \quad (10)$$

*Theorem 3:* A system with dynamics (3) and update law (5) is monotonically convergent above the threshold

$$\kappa = \frac{\|(\mathbf{I} - \mathbf{P}\mathbf{Q}\mathbf{P}^{-1})(\mathbf{r} - \mathbf{d})\|}{1 - \gamma} \quad (11)$$

if

$$\gamma := \|\mathbf{P}\mathbf{Q}(\mathbf{I} - \mathbf{L}\mathbf{P})\mathbf{P}^{-1}\| < 1. \quad (12)$$

*Proof:* See [26]. ∎

## III. PROBLEM FORMULATION

This work considers a multi-agent learning problem, where a set of autonomous agents is given, that is,

$$\mathcal{M} := \{1, \ldots, M\}$$

in which agents are labeled 1 through $M \in \mathbb{N}$ and the $m$th agent's learning dynamics is specified by the respective Q-filter $\mathbf{Q}_m$ and learning gain matrix $\mathbf{L}_m$. All agents are given the same tracking task defined by the reference trajectory $\mathbf{r}$. We distinguish input and output of each agent $m \in \mathcal{M}$ at each trial $j \in \mathbb{N}_0$ by introducing a superscript, i.e., $\mathbf{u}_j^m$ and $\mathbf{y}_j^m$, respectively. The corresponding error trajectory is $\mathbf{e}_j^m$.

*Assumption 1:* All agent are assumed to have the same dynamics (3) defined by plant matrix $\mathbf{P}$ and disturbance $\mathbf{d}$. Thus, the system dynamics of agent $m \in \mathcal{M}$ becomes

$$\forall j \in \mathbb{N}_0, \quad \mathbf{y}_j^m = \mathbf{P}\mathbf{u}_j^m + \mathbf{d}. \quad (13)$$

After each trial, agents are allowed to exchange information, namely their respective input–output, and error trajectories.

*Assumption 2:* The underlying network topology is assumed to be strongly connected, i.e., there exists a path between any ordered pair of agents.

The problem considered in this work consists of designing an ILC-based learning scheme for the MAS that exploits the potential of information exchange and achieves fast, monotonic convergence of the Euclidean norm of all tracking errors to small values.

## IV. COLLECTIVE ILC

The formulated problem can be trivially solved by employing two straightforward strategies that serve as a baseline for the following analysis.

1) *The Single ILC Strategy:* Only one agent learns the task and, after completion, lets other agents copy its solution.
2) *The Parallel ILC Strategy:* All agents learn the task in parallel without sharing any information during the learning process. When the first agent converges to a suitable solution, all other agents copy it. By using agents with different learning gain matrices $\mathbf{L}_m$ and Q-filters $\mathbf{Q}_m$, the chance that at least one agent quickly converges to the solution is increased.

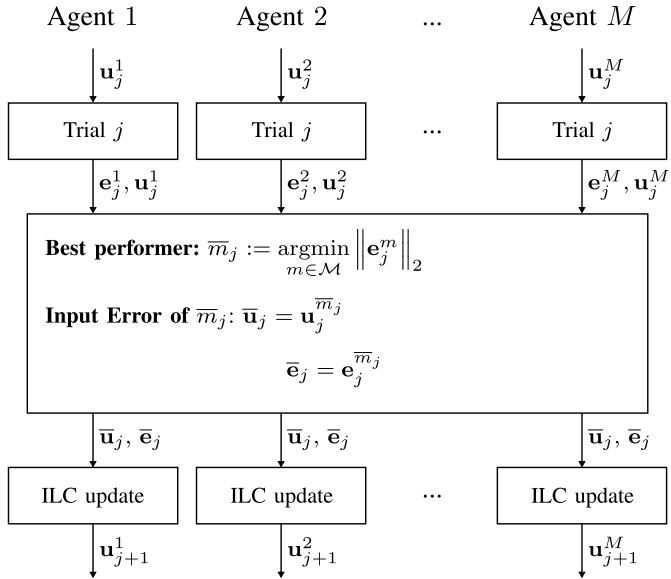

Fig. 2. CILC algorithm for one trial $j$: The best performing agent is determined by comparing the agents' error norms and the best performer's input and error trajectory are used by all agents to perform their respective update law.

As neither of these two strategies stipulates on exchanging information after each trial, the full potential of a multi-agent setting is not exploited. In the proposed CILC strategy, called CILC, agents compare their respective error trajectories after each trial and determine the currently best performing agent. This agent's input and error trajectories are used by all agents to determine their respective next-iteration inputs, as illustrated in Fig. 2. Formally, let agent $\overline{m}_j \in \mathcal{M}$ be the best performer at trial $j$, which is determined by the Euclidean error norm, that is,

$$\forall j \in \mathbb{N}_0, \quad \overline{m}_j := \underset{m \in \mathcal{M}}{\arg\min} \left\| \mathbf{e}_j^m \right\|. \tag{14}$$

Each agent $m \in \mathcal{M}$, instead of using its own current trial information $(\mathbf{u}_j^m, \mathbf{e}_j^m)$ to compute the next input $\mathbf{u}_{j+1}^m$, uses the input and error trajectory copied from the best performer. To this end, let, $\forall j \in \mathbb{N}_0, \overline{\mathbf{u}}_j \in \mathbb{R}^N$ denote the input that agent $\overline{m}_j$ applied at iteration $j$, that is,

$$\overline{\mathbf{u}}_j := \mathbf{u}_j^{\overline{m}_j}. \tag{15}$$

Similarly, let, $\forall j \in \mathbb{N}_0, \overline{\mathbf{e}}_j \in \mathbb{R}^N$ denote the error that the best-performing agent $\overline{m}_j$ achieved at iteration $j$, that is,

$$\overline{\mathbf{e}}_j := \mathbf{e}_j^{\overline{m}_j}. \tag{16}$$

After each trial $j$, the pair $(\overline{\mathbf{u}}_j, \overline{\mathbf{e}}_j)$ is obtained and then used by each agent in its so-called collective update law, that is,

$$\forall j \in \mathbb{N}_0 \quad \forall m \in \mathcal{M}, \quad \mathbf{u}_{j+1}^m = \mathbf{Q}_m(\overline{\mathbf{u}}_j + \mathbf{L}_m \overline{\mathbf{e}}_j). \tag{17}$$

*Remark 2:* Consensus-based leader election approaches allow to execute (14) in a distributed fashion. See [27] for a theoretical analysis. By Assumption 2, agents can compute (14) by solving a leader election problem over a strongly connected network. This requires a number of consensus iterations proportional to the so-called network topology's diameter (see [28]). In general, the communication rate is much faster than any ILC trial dynamics. Hence, the time required by the network to compute (14) is negligible.

*A. Error Dynamics*

To analyze the learning error dynamics, we first combine the system dynamics (13), the best-performer definition (14), and the collective update law (17), which yields

$$\forall j \in \mathbb{N}_0 \quad \forall m \in \mathcal{M} \quad \mathbf{e}_{j+1}^m = \mathbf{\Omega}_m \overline{\mathbf{e}}_j + \mathbf{\Psi}_m (\mathbf{r} - \mathbf{d}) \tag{18}$$

$$\mathbf{\Omega}_m := \mathbf{P} \mathbf{Q}_m (\mathbf{I} - \mathbf{L}_m \mathbf{P}) \mathbf{P}^{-1} \tag{19}$$

$$\mathbf{\Psi}_m := \mathbf{I} - \mathbf{P} \mathbf{Q}_m \mathbf{P}^{-1}. \tag{20}$$

*Definition 4:* The dynamics of the pair $(\overline{\mathbf{u}}_j, \overline{\mathbf{e}}_j)$ from trial to trial is called CILC dynamics.

*Proposition 1:* Given a set $\mathcal{M}$ of agents with dynamics (13) and collective update law (14)–(17), the CILC dynamics is

$$\overline{\mathbf{e}}_{j+1} = \mathbf{\Omega}_{\overline{m}_{j+1}} \overline{\mathbf{e}}_j + \mathbf{\Psi}_{\overline{m}_{j+1}} (\mathbf{r} - \mathbf{d}) \tag{21}$$

$$\overline{\mathbf{u}}_{j+1} = \mathbf{Q}_{\overline{m}_{j+1}} (\overline{\mathbf{u}}_j + \mathbf{L}_{\overline{m}_{j+1}} \overline{\mathbf{e}}_j). \tag{22}$$

*Proof:* The two equations follow directly by combining (15)–(18). ∎

In analogy to single-agent ILC, concepts as stability, residual error, and monotonic error convergence can be extended to the more general case of CILC systems.

*Definition 5 (Monotonic Convergence Above a Threshold in CILC):* A CILC system with dynamics (13) and collective update law (14)–(17) is called monotonically convergent above a threshold $\overline{\kappa}$ under a given norm $\|\cdot\|$ if

$$\forall j \in \mathbb{N}_0, \ \|\overline{\mathbf{e}}_j\| \geq \overline{\kappa} \implies \|\overline{\mathbf{e}}_{j+1}\| \leq \|\overline{\mathbf{e}}_j\|. \tag{23}$$

*Theorem 4:* A CILC system with dynamics (13) and collective update law (14)–(17) is monotonically convergent above a threshold

$$\overline{\kappa} = \frac{\max_{m \in \mathcal{M}} \|\mathbf{\Psi}_m (\mathbf{r} - \mathbf{d})\|}{1 - \overline{\gamma}} \tag{24}$$

where

$$\overline{\gamma} := \max_{\mathbf{v} \in \mathbb{R}^N} \frac{\min_{m \in \mathcal{M}} \|\mathbf{\Omega}_m \mathbf{v}\|}{\|\mathbf{v}\|} \tag{25}$$

is the collective convergence rate, if

$$\overline{\gamma} < 1. \tag{26}$$

*Proof:* Combining the error dynamics (18) and collective strategy (14) yields

$$\|\overline{\mathbf{e}}_{j+1}\| = \min_{m \in \mathcal{M}} \|\mathbf{\Omega}_m \overline{\mathbf{e}}_j + \mathbf{\Psi}_m (\mathbf{r} - \mathbf{d})\| \tag{27}$$

which, by applying the inequality of norms and min-max operators, leads to

$$\|\overline{\mathbf{e}}_{j+1}\| \leq \min_{m \in \mathcal{M}} \|\mathbf{\Omega}_m \overline{\mathbf{e}}_j\| + \max_{m \in \mathcal{M}} \|\mathbf{\Psi}_m (\mathbf{r} - \mathbf{d})\|. \tag{28}$$

Note that by (25), for any $\overline{\mathbf{e}}_j$

$$\min_{m \in \mathcal{M}} \|\mathbf{\Omega}_m \overline{\mathbf{e}}_j\| \leq \overline{\gamma} \|\overline{\mathbf{e}}_j\|. \tag{29}$$

Combining the latter with (28) and substracting $\|\overline{\mathbf{e}}_j\|$ gives

$$\|\overline{\mathbf{e}}_{j+1}\| - \|\overline{\mathbf{e}}_j\| \leq (\overline{\gamma} - 1)\|\overline{\mathbf{e}}_j\| + \max_{m \in \mathcal{M}} \|\mathbf{\Psi}_m (\mathbf{r} - \mathbf{d})\|. \tag{30}$$

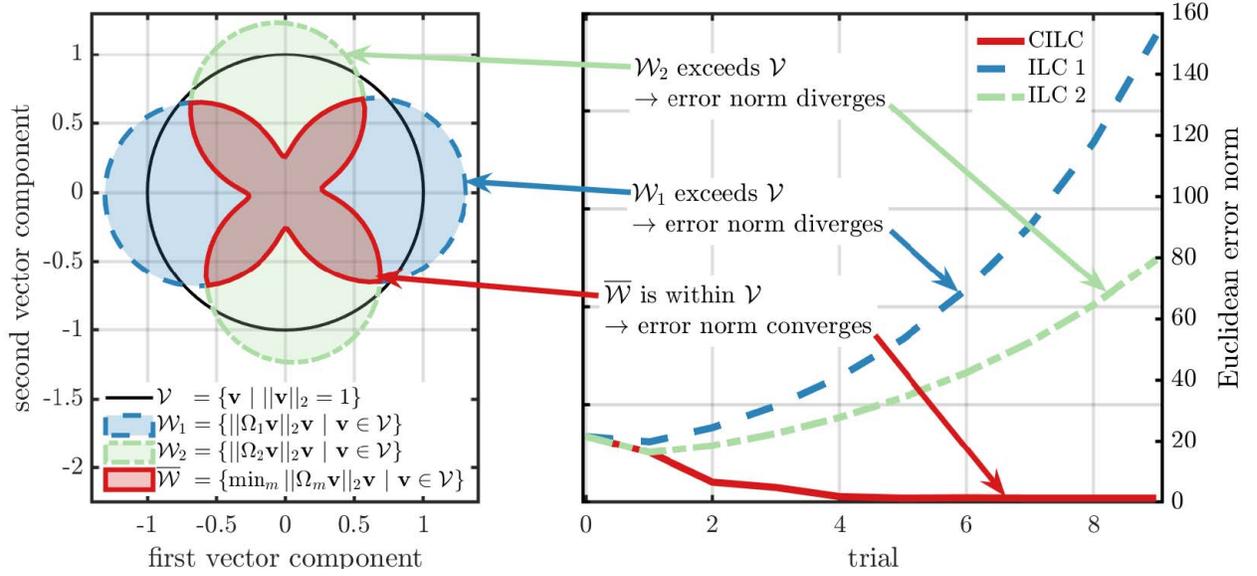

Fig. 3. (Left) Graphical presentation of the monotonic convergence conditions for single-agent and CILC in an example with two agents: If the affine maps of the unit circle remain within the unit circle, the respective system is monotonically convergent above a threshold. (Right) Error norm progressions of the isolated agents and the corresponding CILC in one simulation example: In the single-agent ILC setup, the error norms of both agents diverge. Using the same agents in the CILC setup, the error norm converges monotonically.

By definition (23), monotonic convergence is given if

$$\|\bar{\mathbf{e}}_{j+1}\| - \|\bar{\mathbf{e}}_j\| \leq 0 \tag{31}$$

which, according to (30), is implied by

$$(\bar{\gamma} - 1)\|\bar{\mathbf{e}}_j\| + \max_{m \in \mathcal{M}} \|\mathbf{\Psi}_m(\mathbf{r} - \mathbf{d})\| \leq 0 \tag{32}$$

equivalently

$$\|\bar{\mathbf{e}}_j\| \geq \frac{\max_{m \in \mathcal{M}} \|\mathbf{\Psi}_m(\mathbf{r} - \mathbf{d})\|}{1 - \bar{\gamma}}. \tag{33}$$

Monotonic convergence, i.e., (31), is implied by (33). This concludes the proof. ∎

For CILC to be monotonically convergent above a threshold, Theorem 4 does not require the individual agents to be monotonically convergent themselves. Thus, the proposed collective approach allows a group of agents to obtain a property that none of the individual agents has. To illustrate this fact, we consider a scenario with two agents with plant matrix (49), Q-filter matrices (50), and learning gain matrices (50) given in the appendix. By Theorems 2 and 3, single-agent monotonic convergence under the Euclidean norm is given if the respective $\mathbf{\Omega}_m$ is a contraction. Geometrically, this means that $\mathbf{\Omega}_m$ maps all points on the $N$-dimensional unit sphere

$$\mathcal{V} = \{\mathbf{v} \in \mathbb{R}^N \mid \|\mathbf{v}\|_2 = 1\}$$

into that sphere. In the given example, $N = 2$ is chosen, and

$$\mathcal{W}_m := \{\|\mathbf{\Omega}_m \mathbf{v}\|_2 \mathbf{v} \mid \mathbf{v} \in \mathcal{V}\}$$

is plotted in Fig. 3. The plot shows that neither of the individual ILC systems is monotonically convergent, since both $\mathcal{W}_1$ and $\mathcal{W}_2$ partially exceed $\mathcal{V}$. Formally, this is due to the fact that $\mathbf{\Omega}_1$ and $\mathbf{\Omega}_2$ are not contractions, that is

$$\exists \mathbf{v} \in \mathcal{V}, \quad \|\mathbf{\Omega}_i \mathbf{v}\|_2 > 1 \quad i \in \{1, 2\}.$$

However, combining $\mathbf{\Omega}_1$ and $\mathbf{\Omega}_2$ by the CILC approach is beneficial. In fact, one can see from Fig. 3 that

$$\overline{\mathcal{W}} := \{\min_{m \in \mathcal{M}} \|\mathbf{\Omega}_m \mathbf{v}\|_2 \mathbf{v} \mid \mathbf{v} \in \mathcal{V}\}$$

lies completely within $\mathcal{V}$. This implies that, $\forall \mathbf{v} \in \mathcal{V}$

$$\|\min_{m \in \mathcal{M}} \|\mathbf{\Omega}_m \mathbf{v}\|_2 \mathbf{v}\| = \min_{m \in \mathcal{M}} \|\mathbf{\Omega}_m \mathbf{v}\|_2 \|\mathbf{v}\|_2 < \|\mathbf{v}\|_2$$

equivalently, $\forall \mathbf{v} \in \mathcal{V}$, $\min_{m \in \mathcal{M}} \|\mathbf{\Omega}_m \mathbf{v}\|_2 < 1$, thus verifying (26). This means that the CILC is monotonically convergent above a threshold, although none of the individual ILC systems is monotonically convergent. This finding is confirmed and further illustrated by simulating the evolution of error norms through trials (see Fig. 3). In fact, both individual ILCs exhibit a diverging error norm, whereas the error of the CILC converges monotonically.

We now analyze the properties of CILC systems in which at least one of the individual ILCs is monotonically convergent.

*Theorem 5:* A CILC system with dynamics (13) and collective update law (14)–(17) is monotonically convergent above a threshold $\bar{\kappa}$ under a given norm $\|\cdot\|$ if at least one agent $m \in \mathcal{M}$ is monotonically convergent above the same threshold.

*Proof:* Agent $p \in \mathcal{M}$ is monotonically convergent above threshold $\bar{\kappa}$, namely, $\forall j \in \mathbb{N}_0$

$$\|\mathbf{e}_j^p\| \geq \bar{\kappa} \implies \|\mathbf{e}_{j+1}^p\| \leq \|\mathbf{e}_j^p\|.$$

By (18) and since $p$ is monotonically convergent above $\bar{\kappa}$

$$\|\bar{\mathbf{e}}_j\| \geq \bar{\kappa} \implies \|\mathbf{e}_{j+1}^p\| \leq \|\bar{\mathbf{e}}_j\|.$$

By the collective update law (14)–(17), $\forall j \in \mathbb{N}_0$

$$\|\bar{\mathbf{e}}_{j+1}\| \leq \|\mathbf{e}^p_{j+1}\|.$$

By bringing the latter two equations together, one obtains, $\forall j \in \mathbb{N}_0$

$$\|\bar{\mathbf{e}}_j\| \geq \bar{\kappa} \implies \|\bar{\mathbf{e}}_{j+1}\| \leq \|\bar{\mathbf{e}}_j\|$$

which is equivalent to (23), thus the proof is concluded. ∎

*Corollary 1:* The considered CILC system is monotonically convergent above the threshold determined by the smallest threshold of all individual ILCs that are monotonically convergent above a threshold.

*Proof:* The proof follows directly from the proof of Theorem 5, by considering $\bar{\kappa} := \arg\min_{m \in \mathcal{M}}(\kappa_p)$. ∎

*Remark 3:* The extension of Theorem 5 to the case of uncertain plant dynamics is straightforward if one of the agents guarantees robust monotonic convergence. Hence, the robust design of one agent guarantees robust monotonic convergence of the collective.

*Definition 6 (Asymptotic Stability of CILC):* A CILC system with dynamics (13) and collective update law (14)–(17) is called asymptotically stable if the limit

$$\lim_{j \to \infty} \bar{\mathbf{e}}_j = \bar{\mathbf{e}}_R \quad (34)$$

exists and is unique.

*Theorem 6:* The considered CILC system is asymptotically stable if one agent $m \in \mathcal{M}$ is monotonically convergent above a threshold with residual error $\mathbf{e}_R^m = \mathbf{0}$.

*Proof:* By (8) and since $\mathbf{e}_R^m = \mathbf{0}$, the individual ILC has

$$\forall j \in \mathbb{N}_0, \quad \|\mathbf{e}^m_{j+1}\| \leq \gamma_m \|\mathbf{e}^m_j\| \quad (35)$$

with $\gamma_m < 1$. By the latter and CILC error dynamics (18)

$$\forall j \in \mathbb{N}_0, \quad \|\mathbf{e}^m_{j+1}\| \leq \gamma_m \|\bar{\mathbf{e}}_j\|.$$

By the collective update law (14)-(17), $\forall p \in \mathcal{M}$

$$\|\bar{\mathbf{e}}_{j+1}\| \leq \|\mathbf{e}^p_{j+1}\|$$

which yields

$$\forall j \in \mathbb{N}_0, \quad \|\bar{\mathbf{e}}_{j+1}\| \leq \gamma_m \|\bar{\mathbf{e}}_j\|.$$

The latter trivially verifies

$$0 \leq \lim_{j \to \infty} \|\bar{\mathbf{e}}_j\| \leq \lim_{j \to \infty} \gamma_m^j \|\bar{\mathbf{e}}_0\|.$$

With $\gamma_m < 1$, the proof is concluded. ∎

*Remark 4:* In the case of trial-invariant, single-agent ILC with linear dynamics, monotonic convergence is a stronger property than asymptotic stability, meaning that the former implies the latter. This is not the case for CILC systems, for which the learning dynamics (21) can be trial-variant. A CILC system may be monotonically convergent above a threshold while not being asymptotically stable, as further illustrated in the following simulation studies.

*Remark 5:* The analysis of this section has focused on the properties of asymptotic stability and monotonic convergence. A theoretical analysis of the CILC's performance in comparison to that of single-agent ILC for a limited number of trials is given in Appendix B.

*Remark 6 (Communication Network Topology):* In manufacturing, industry, and many other applications, the implemented network topology is fully connected. In Remark 2, however, we pointed out that the collective update law (17) can be obtained by all agents at every iteration by assuming a strongly connected underlying network topology. This is a consistent relaxation of requiring a fully connected topology. If the network topology is disconnected, one can immediately verify that any strongly connected subgraph containing a monotonically convergent agent will also be monotonically convergent.

### B. CILC Design

Section IV-A has shown that cooperation can lead to advantages in multi-agent ILC. Now, we consider the application of the norm-optimal ILC (NO-ILC) framework to CILC such that collective learning performance is maximized. The Q-filter $\mathbf{Q}_m$ and learning matrix $\mathbf{L}_m$ of the $m$th agent are chosen to minimize the next-trial cost criterion

$$J := \|\mathbf{e}_{j+1}\|_2^2 + s_m \|\mathbf{u}_{j+1} - \mathbf{u}_j\|_2^2 + r_m \|\mathbf{u}_{j+1}\|_2^2 \quad (36)$$

where $s_m \in \mathbb{R}_{>0}$ and $r_m \in \mathbb{R}_{>0}$ are positive weights with the following effects [29]:

1) raising $s_m$ increases robustness but slows convergence;
2) raising $r_m$ increases robustness but also residual errors.

In single-agent ILC, $s_m$ and $r_m$ typically are chosen such that a reasonable trade-off between robustness, small residual errors, and rapid convergence is achieved (see [30]). However, CILC can overcome these limitations by designing the respective agents such that each of them guarantees one of the desired properties. For example, if we are interested in achieving robust monotonic convergence of the collective, one of the agents should be designed by choosing comparatively large values for $s_m$ and $r_m$ such that this agent guarantees robust monotonic convergence. According to Theorem 5, monotonic convergence of this one agent is a sufficient condition for the collective error norm to also monotonically converge. Hence, the robustness of one agent also guarantees robust convergence of the collective. Similarly, if we are interested in converging to a residual error of zero, one agent should be designed using a comparatively large value for $s_m$ but a value of zero for $r_m$ such that this agent monotonically converges to a residual error of zero. According to Theorem 6, this guarantees that the collective error norm also converges to zero. If one is interested in increasing the speed of convergence, one of the agents should be designed with a comparatively small value of $s_m$. This choice can lead to a rapid decline of the error norm on initial trials but may come at the price of error divergence on later trials if the configuration was employed in single-agent ILC. However, if such an agent is combined with one that guarantees robust monotonic convergence, the collective error norm is not only expected to initially decline at a rapid rate but to also continue declining due to the monotonic convergence of the robust agent. In summary, we propose that the individual agents should be designed heterogeneously such that each of them guarantees one of the desired properties. The following simulation studies aim at verifying that CILC can merge the advantages of the individual agents.

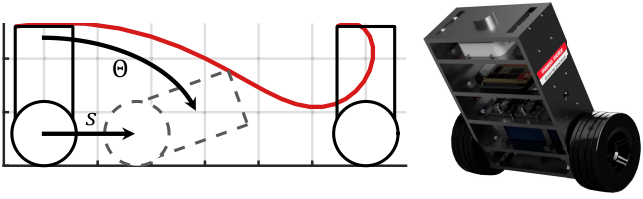

Fig. 4. Sketch of the maneuver considered in simulation (left) and a rendering of a TWIPR (right).

## V. SIMULATIONAL AND EXPERIMENTAL VALIDATION

To investigate how the design of individual agents' learning laws affects the performance of CILC compared to that of the isolated agents, we first perform simulation analyses. We consider two typical performance criteria for ILC systems, namely 1) small residual errors and 2) fast convergence to those values. As discussed in the previous section, model uncertainty makes it generally difficult to design a single agent that robustly satisfies both performance criteria simultaneously. However, it is typically far less difficult to design an ILC agent that achieves either one of the criteria. Hence, we propose a CILC design that assigns complementary advantages and disadvantages to different agents, such that the collective learning dynamics exhibit a combination of the respective advantages and simultaneously satisfies all of the performance criteria. In the following, we validate this strategy and its practical applicability in simulational and experimental studies of a robotic MAS.

### A. Testbed

We consider a group of TWIPRs, supposed to perform repeated maneuvers. The same robot has previously been used for validation of single-agent ILC methods (see [31]). Each robot consists of the pendulum body housing the main electronics including a microcomputer, inertial measurement units, motors, and battery. Wheels are mounted onto the motor such that the robot can drive while balancing the chassis. The system is stabilized by feedback control, which is designed based on an approximate plant model with uncertain parameters.

First, let us briefly introduce the dynamics of a TWIPR moving along a straight line. The motor torque is the input variable denoted by $u \in \mathbb{R}$. Let $\Theta \in \mathbb{R}$ denote the pendulum's pitch angle and let $s \in \mathbb{R}$ denote the robot's position (see Fig. 4). The state vector is defined as

$$\mathbf{z} = [\theta \ \dot{\theta} \ s \ \dot{s}]^T. \quad (37)$$

A thorough derivation of the TWIPR's dynamics can be found in [32]. By this, let the dynamics be

$$\forall t \in \mathbb{R}_{\geq 0}, \quad \dot{\mathbf{z}}(t) = f(\mathbf{z}(t), u(t)). \quad (38)$$

This latter is discretized and linearized at the upright equilibrium, thus obtaining the dynamics

$$\forall n \in \mathbb{N}_0, \quad \mathbf{z}(n+1) = \mathbf{A}\mathbf{z}(n) + \mathbf{B}u(n). \quad (39)$$

To stabilize the inverted pendulum, the controller input $u_C \in \mathbb{R}$ is calculated by a discrete-time feedback controller of the form

$$\forall n \in \mathbb{N}_0, \quad u_C(n) = -\mathbf{K}\mathbf{z} \quad (40)$$

with sampling period $T = 0.02$ s. The feedback matrix $\mathbf{K} \in \mathbb{R}^{1 \times 4}$ is designed using pole placement, a conservative choice of poles, and the linearized dynamics (39).

This feedback controller successfully stabilizes the upper equilibrium but does not achieve accurate reference tracking for agile maneuvers. It has been demonstrated that adding an ILC enables the robotic agent to learn such maneuvers through repeated trials [31]. As demonstrative example, the TWIPR is meant to perform the pitch angle trajectory

$$\forall n \in \{0, \ldots, 99\}, \quad r(n) = 30\sin(\pi T n) \ [°] \quad (41)$$

for which the feedback controller does not achieve sufficiently precise tracking performance and hence is complemented by an ILC that computes the input $u_{\text{ILC}} \in \mathbb{R}$ leading to the overall system input

$$\forall n \in \mathbb{N}_0, \quad u(n) = u_C(n) + u_{\text{ILC}}(n) \quad (42)$$

and time-domain dynamics

$$\forall n \in \mathbb{N}_0, \quad \mathbf{z}(n+1) = (\mathbf{A} - \mathbf{B}\mathbf{K})\mathbf{z}(n) + \mathbf{B}u_{\text{ILC}}(n). \quad (43)$$

For designing and analyzing the ILC, the lifted form dynamics (3) are used, where $\mathbf{P}$ is a lower triangular, Toeplitz matrix of the Markov parameters

$$\forall i \in \mathbb{N}, \quad p_i = \mathbf{C}\sum_{k=1}^{i}(\mathbf{A}-\mathbf{BK})^{i-1-k}\mathbf{B} \quad (44)$$

such that

$$\mathbf{P} = \begin{bmatrix} p_1 & 0 & 0 & \cdots & 0 \\ p_2 & p_1 & 0 & \cdots & 0 \\ p_3 & p_2 & p_1 & \cdots & 0 \\ \vdots & \vdots & \vdots & \ddots & \vdots \\ p_N & p_{N-1} & p_{N-2} & \cdots & p_1 \end{bmatrix}. \quad (45)$$

We consider the case in which several agents should simultaneously learn to perform a given maneuver accurately in as few trials as possible. While it appears obvious that the aforementioned baseline strategies might be used to solve this task within some potentially large but finite number of trials, we aim at investigating to which extent the proposed CILC method improves the learning performance of the MAS.

The Q-filter $\mathbf{Q}_m$ and learning matrix $\mathbf{L}_m$ of the $m$th agent are determined by NO-ILC, as described in Section IV-B.

### B. Simulation

The nonlinear dynamics (38) are used to simulate the actual robot dynamics. To simulate model uncertainty, the NO-ILC design is based on the linearized dynamics with plant matrix $\mathbf{P}$, where the inertia parameters of the robot have been increased by 40% to simulate model uncertainty. Three agents are considered, and a different ILC law is designed for each of them.

1) The greedy agent (ILC G) is designed assuming precise model information meaning that it aims at achieving

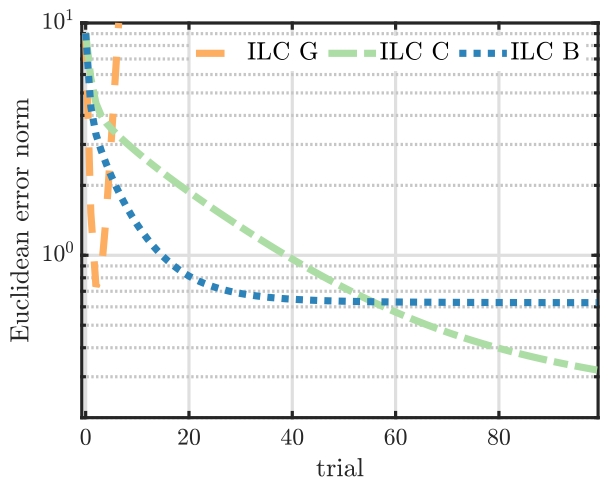

Fig. 5. Performance of the individual agents; ILC G: greedy agent with a rapid initial decline in error norm but unstable dynamics; ILC C: conservative agent with smallest residual error, monotonic convergence but slow speed of convergence; ILC B: balanced agent with decent residual error norm, decent speed of convergence, and monotonic convergence.

   small residual errors and a fast speed of convergence but neglects the need for robustness.
2) The conservative agent (ILC C) targets robust convergence to zero residual errors, which comes at the price of a slow speed of convergence.
3) The balanced agent (ILC B) is designed to assure robustness while aiming at a trade-off between small residual errors and a fast speed of convergence.

The characteristics of the individual agents are verified by simulation with the progressions of the respective error norms depicted in Fig. 5. Note that the conservative and greedy agents are based on rather extreme design choices, which require little tuning and would typically not be employed in single-agent ILC. The balanced agent is a realistic design decision for single-agent ILC, which requires comparatively more tuning to achieve a satisfying trade-off between the performance criteria.

To investigate the rich effects of combining agents with various characteristics using CILC, we consider four different MASs, each of which is formed by choosing a different subset of these agents.

Conservative + Balanced agent (see Fig. 6): When combining the conservative and balanced agent using CILC, the error norm initially declines at the faster pace of the balanced agent. After reaching the residual error of the balanced agent, the error norm of CILC continues to decline at the rate of the conservative agent converging to its smaller residual error. Note that in this constellation CILC achieves an error norm smaller or equal to that of either isolated agent on each trial. Hence, the CILC combines the individual advantages of the involved agents, namely the faster initial convergence rate of the balanced agent and the smaller residual error of the conservative agent.

Conservative + greedy agent (see Fig. 6): When combining the conservative and greedy agent using CILC, the error norm initially declines at the rapid pace of the greedy agent. Once the greedy agent's error norm starts to diverge, the CILC's error norm continues to decline at the pace of the conservative agent, again, converging to its residual error. Note that the CILC combines the advantageous properties of the involved agents even if their isolated dynamics are diverging.

Balanced + greedy agent (see Fig. 6): When combining the balanced and greedy agent, the learning process consists of three distinct phases. In the first phase, the CILC's error norm declines at the rapid pace of the greedy agent, which is the sole best performer over these trials. The second phase begins when the greedy agent's error norm starts to diverge. Here, neither of the two agents is capable of further decreasing the error norm, and the agent leading to the smallest increase in error norm, namely the balanced agent, is the sole best performer. The third phase starts when the CILC's error norm reaches the residual error norm of the balanced agent. Here, the CILC's error norm oscillates, which is caused by a periodic switching between the greedy and balanced agent as the best performer. Therefore, the learning dynamics are trial-varying, and the CILC is not asymptotically stable despite being monotonically convergent above a threshold, which verifies Remark 4. The issue of oscillating error norms can be circumvented by adjusting the collective learning law such that the input trajectories are only updated if at least one of the agents is capable of decreasing the error norm.

Conservative + balanced + greedy agent (see Fig. 6): Lastly, the collective of all three agents is considered. Again, the CILC's error norm initially declines at the rapid pace of the greedy agent and continues to decline to the zero residual error of the conservative agent. Comparison of this collective's error norm progression with the one of combining just the conservative and greedy agent shows that further including the balanced agent does not lead to a noticeable increase in performance for this specific scenario (see Fig. 7).

In summary, the following statements can be made regarding the simulations.
1) CILC is capable of improving learning performance by merging individual agents' advantages, as, for example, rapid (initial) convergence, small residual errors, or robustness.
2) The performance increase achieved by cooperation is the larger, the more the agents' characteristics differ such that they ideally have complementary advantages.
3) Adding agents to the collective that bring no additional advantages for the specific scenario does neither lead to an increase nor to a decrease in performance.

### C. Experiment

To further validate the practical applicability of CILC, the method is applied in experiments using two TWIPRs depicted in Fig. 8. As in the simulation study, we consider three different MASs, each of which consists of agents with different characteristics. The first experiment employs two monotonically convergent agents with different speeds of convergence and residual error norms. In the second experiment, a monotonically convergent agent is combined with an agent whose learning dynamics are unstable for the given scenario. And, in the third experiment, we consider the case of two monotonically convergent agents with one of the two achieving a smaller error norm than the other on each trial. The agents are designed using NO-ILC and uncertain estimates of the linearized plant dynamics.

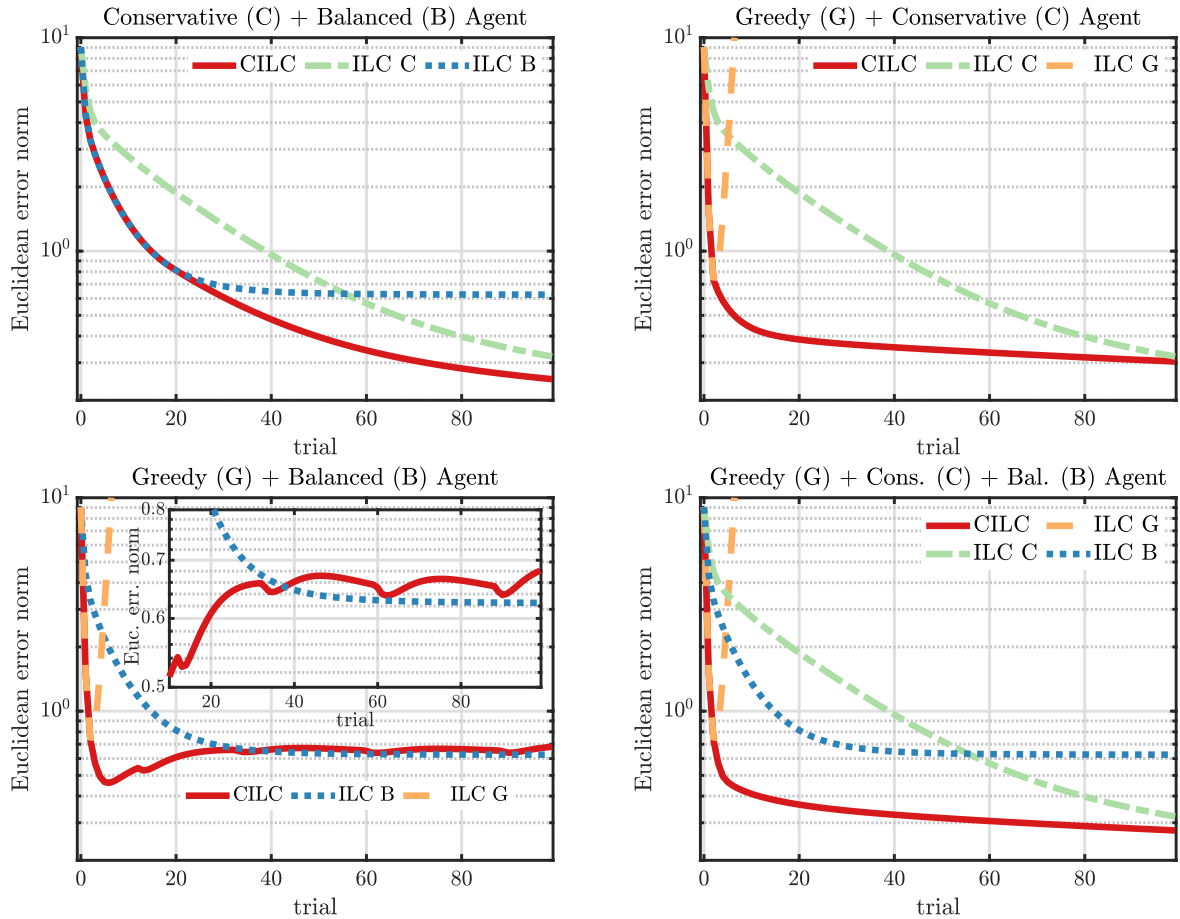

Fig. 6. Comparison of the isolated ILC systems' and CILC's error norm progression with four different collectives formed from the three agents. The CILC is capable of increasing performance compared to the isolated agents by merging the advantages of the respective agents such as a fast speed of convergence, small residual errors, and monotonic convergence above a threshold.

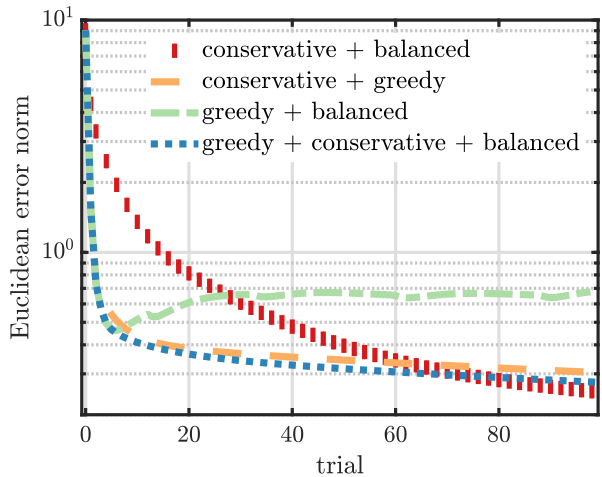

Fig. 7. Progression of the CILC collectives' error norms: Combining agents with complementary advantageous, e.g., greedy + conservative, leads to better performance than combining agents with similar characteristics, e.g., conservative + balanced.

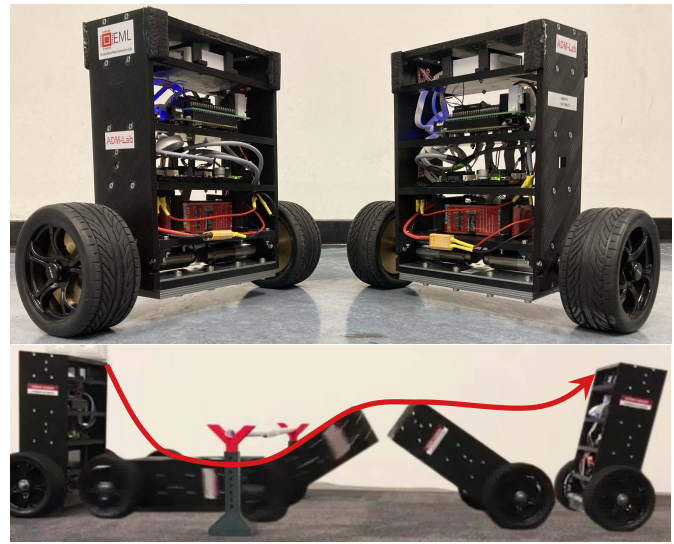

Fig. 8. (Top) Two TWIPRs form the robotic collective that is employed in experiment. (Bottom) The collective has to learn a maneuver that is similar to diving beneath an obstacle.

*Experiment 1:* In the first experiment, ILC 1 is designed to converge at a slower rate but to a smaller residual error norm than ILC 2, while both are designed to be reasonably robust. The specific weights are

$$s_1 = 5, \quad r_1 = 0.1, \quad s_2 = 0.05, \quad r_2 = 1. \tag{46}$$

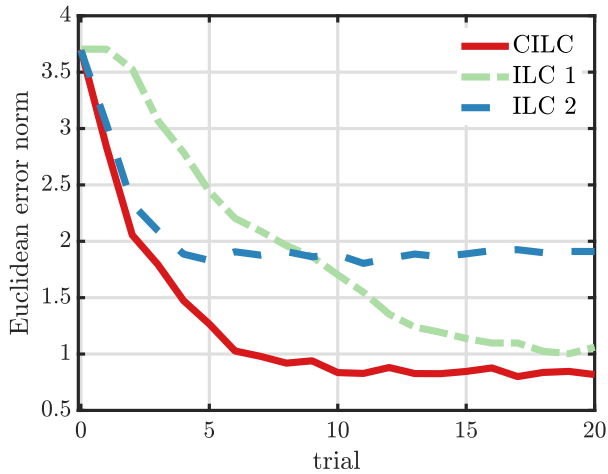

Fig. 9. Progression of the Euclidean error norms in Experiment 1: CILC merges the complementary advantages of the respective agents, namely the small residual error norm of ILC 1 and the fast initial convergence of ILC 2.

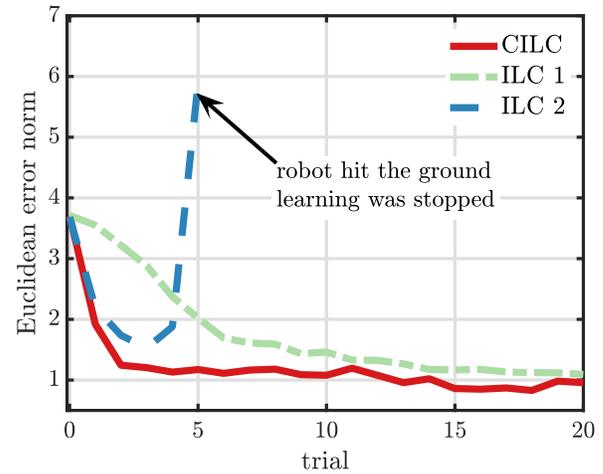

Fig. 10. Progression of the Euclidean error norms in Experiment 2: CILC merges the complementary advantages of the respective agents, namely the monotonic convergence of ILC 1 and the fast initial convergence of ILC 2.

The isolated-ILC results depicted in Fig. 9 show that ILC 1 indeed converges at a slower rate than ILC 2 but achieves a smaller error norm from the seventh trial onward. As in the simulation results, the CILC combines the advantages of the isolated ILC systems, meaning that its error norm initially declines at a rate similar to ILC 2, but continues to decline, similar to ILC 1, afterward. Overall, the CILC achieves an error norm equal or smaller than either of the isolated ILC systems on each trial.

*Experiment 2:* In the second experiment, ILC 1 is the same as before, while ILC 2 is designed to converge at a clearly faster rate than ILC 1 but at the cost of poor robustness. The design parameters are

$$s_1 = 5, \quad r_1 = 0.1, \quad s_2 = 0.005, \quad r_2 = 0.001. \quad (47)$$

The results depicted in Fig. 10 show that the isolated error dynamics of ILC 2 starts to diverge from the third trial onward such that the TWIPR hits the ground on the sixth trial and the learning had to be stopped. The error norm of ILC 1 decreases much slower on the first 3–5 trials, but it is monotonically decreasing over all trials. As before, the MAS with the proposed CILC unites the advantages of the involved agents. The error norm initially declines at the rapid pace of ILC 2 and afterward converges to a small residual error. Overall, the CILC achieves an error norm smaller or equal than either of the isolated ILC systems on each trial.

*Experiment 3:* Lastly, we design ILC 2 to achieve strictly better performance than ILC 1. The design parameters are

$$s_1 = 5, \quad r_1 = 0.1, \quad s_2 = 0.5, \quad r_2 = 0.01. \quad (48)$$

The results depicted in Fig. 11 show that the error norm of ILC 2 not only converges at a significantly faster rate than ILC 1 but also achieves strictly better performance across all trials. As shown in the plot, the error norm of the MAS employing CILC decays at the same rate as the error of isolated ILC 2, meaning that the CILC performance does not decrease when an agent with strictly worse performance is added to the collective.

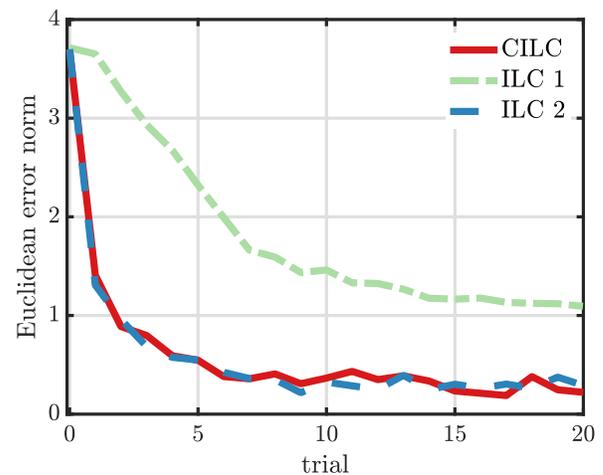

Fig. 11. Progression of the Euclidean error norms in Experiment 3: A CILC system that combines two agents, one of which has strictly worse performance (ILC 1) than the other (ILC 2), performs as good as the strictly better agent. This exemplifies that adding agents with bad isolated performance to a MAS with CILC does not impair the collective performance.

In summary, the experiments confirm the findings of the simulation study. CILC achieves improved learning performance by merging the advantages of the respective agents. This effect can be exploited by designing the involved agents such that they have complementary properties and advantages. Even agents with design choices that would be risky and unreasonable in single-agent ILC can be added to the collective without jeopardizing the performance of the proposed CILC method. Lastly, it should be noted that, due to manufacturing tolerances, the dynamics of the two robots slightly differ such that Assumption 1 is violated. However, the experiments confirm that the proposed CILC nonetheless merges the advantages of the respective agents to increase learning performance.

## VI. CONCLUSION

We have proposed a collective learning method for solving independent-action tasks in which multiple agents must learn to track an identical reference trajectory. Each agent is assigned a different learning law. A collective update law merges the advantages of all individual agents, such as fast initial learning progress, monotonic convergence, and small residual errors. Thereby, the collective approach overcomes limitations that are insuperable in single-agent ILC.

In the theoretical analysis, the standard concepts of asymptotic stability and monotonic convergence were extended to the multi-agent case, and we proved, among others, that the proposed CILC can guarantee monotonic convergence even if none of the individual agents is monotonically convergent. All theoretical findings were validated by simulation and by experiments in a robotic testbed.

The proposed method is expected to be useful for a range of applications in which MASs must solve problems that contain independent-action tasks. CILC allows the group to exploit its heterogeneity and to solve a given reference tracking problem efficiently even in cases in which none of the individuals would succeed on its own. This shows that leveraging the combination of diversity and knowledge transfer in MAS is advantageous even in situations in which collaboration is not a requirement.

Ongoing and future work aims at extending the proposed method to a wider class of MASs, including the case of agent-individual dynamics and prior knowledge. Further research will focus on relaxing the assumption of a strongly connected network topology and analyzing convergence properties under switching topologies. Finally, we will investigate the synergistic potential of heterogeneity and knowledge transfer for closely related multi-agent tasks such as exploration or motion planning.

## APPENDIX A
## SIMULATION EXAMPLE

For the example given in Section IV, the plant matrix is given by

$$\mathbf{P} = \begin{bmatrix} 1 & 0 \\ 0.25 & 1 \end{bmatrix} \qquad (49)$$

the Q-filter matrices are given by

$$\mathbf{Q}_1 = \begin{bmatrix} 1 & 0 \\ 0.1 & 0.2 \end{bmatrix} \quad \mathbf{Q}_2 = \begin{bmatrix} 0.25 & 0 \\ -0.1 & 1.15 \end{bmatrix} \qquad (50)$$

and the learning matrices are given by

$$\mathbf{L}_1 = \begin{bmatrix} -0.3 & 0 \\ 0 & -0.3 \end{bmatrix} \quad \mathbf{L}_2 = \begin{bmatrix} -0.07 & 0 \\ 0.02 & -0.07 \end{bmatrix}. \qquad (51)$$

## APPENDIX B
## PERFORMANCE EVALUATION

Section IV considers static performance (i.e., converging properties) of the CILC system. In this section, we provide some mathematical tools for comparing the dynamic performance (i.e., transient) of CILC and isolated ILCs. By isolated ILC we denote a standard ILC-based system [see (3)] that does not exchange any type of information with other agents. To this end, let

$$\forall j \in \mathbb{N}_0, \quad \tilde{\mathbf{e}}_j^m$$

TABLE I
ERROR VARIABLES

| Error | Meaning |
|---|---|
| $\mathbf{e}_j^m$ | CILC: error of the collaborative individual ILC $m \in \mathcal{M}$ at iteration $j$. |
| $\overline{\mathbf{e}}_j$ | CILC: error of the best performer within the CILC at iteration $j$. |
| $\tilde{\mathbf{e}}_j^m$ | ILC: error of an isolated ILC mounted on agent $m$ at iteration $j$. |

denote the error of the isolated ILC of agent $m \in \mathcal{M}$ at its $j$th iteration. In order to avoid confusion between the three defined errors, we summarize them in Table I. A CILC that outperforms, in terms of dynamic performance, all individual isolated ILCs is called well-performing.

*Definition 7:* Let $\mathbf{e}_0 \in \mathbb{R}^N$ be the initial error trajectory (for both the CILC and all the ILCs). The considered CILC is said to be well-performing if

$$\forall \mathbf{e}_0 \in \mathbb{R}^N \quad \forall (\mathbf{r} - \mathbf{d}) \in \mathbb{R}^N \quad \forall j \in \mathbb{N}_0, \quad \|\overline{\mathbf{e}}_j\|_2 \leq \|\tilde{\mathbf{e}}_j^m\|_2,$$
$$m = 1, \ldots, M.$$

However, the CILC system can be well performing only under particular conditions.

*Definition 8:* The considered CILC is said to be *well-performing for* the pair $(\mathbf{e}_0, \mathbf{r} - \mathbf{d})$ with $\mathbf{e}_0 \in \mathbb{R}^N$, $(\mathbf{r} - \mathbf{d}) \in \mathbb{R}^N$, if

$$\forall j \in \mathbb{N}_0, \quad \|\overline{\mathbf{e}}_j\|_2 \leq \|\tilde{\mathbf{e}}_j^m\|, \quad m = 1, \ldots, M. \qquad (52)$$

At this point, one can obtain the error dynamics for the case of individual isolated ILCs (i.e., $\tilde{\mathbf{e}}_j^m$) and for the case of a CILC (i.e., $\overline{\mathbf{e}}_j$), as function of the initial error trajectory $\mathbf{e}_0$ and the vector $(\mathbf{r} - \mathbf{d})$.

*Proposition 2 (Isolated ILC error):* The error of the individual isolated ILC $m \in \mathcal{M}$ is, $\forall j \in \mathbb{N}_0$

$$\tilde{\mathbf{e}}_j^m = \tilde{\mathbf{A}}_m^j \mathbf{e}_0 + \tilde{\mathbf{B}}_m^j (\mathbf{r} - \mathbf{d}) \qquad (53)$$

where

$$\tilde{\mathbf{A}}_m^j := (\mathbf{\Omega}_m)^j$$

and

$$\tilde{\mathbf{B}}_m^j := \sum_{p=1}^{j} (\mathbf{\Omega}_m)^{j-p} \mathbf{\Psi}_m.$$

*Proof:* Analog to (18), the error dynamics of the individual isolated ILC $m \in \mathcal{M}$ is, $\forall j \in \mathbb{N}_0$

$$\tilde{\mathbf{e}}_{j+1}^m = \mathbf{\Omega}_m \tilde{\mathbf{e}}_j^m + \mathbf{\Psi}_m (\mathbf{r} - \mathbf{d})$$

where $\mathbf{\Omega}_m$ and $\mathbf{\Psi}_m$ are defined in, respectively, (19) and (20). The error dynamics can be recursively expanded, until (53) is obtained. This concludes the proof. ∎

*Proposition 3 (CILC Error):* The error of the CILC system is, $\forall j \in \mathbb{N}_0$

$$\overline{\mathbf{e}}_j = \overline{\mathbf{A}}^j \mathbf{e}_0 + \overline{\mathbf{B}}^j (\mathbf{r} - \mathbf{d}) \qquad (54)$$

where

$$\bar{\mathbf{A}}^j := \mathbf{\Omega}_{f_j}\mathbf{\Omega}_{f_{j-1}}\cdot\ldots\cdot\mathbf{\Omega}_{f_1}\mathbf{\Omega}_{f_0} = \prod_{i=0}^{j}\mathbf{\Omega}_{f_{j-i}} \quad (55)$$

and

$$\bar{\mathbf{B}}^j := \sum_{p=1}^{j}\left(\prod_{l=0}^{j-1-p}\mathbf{\Omega}_{f_{j-l}}\right)\mathbf{\Psi}_{f_p}. \quad (56)$$

*Proof:* By recursively expanding (21), one obtains (54). The proof is concluded. ∎

Besides considering these two scenarios, i.e., the isolated ILCs and the CILC system as a whole, we need to derive the error dynamics for the individual ILC as a component of the CILC system, also referred to as collaborative individual ILC.

*Proposition 4 (Collaborative ILC Error):* The error of the collaborative ILC within a CILC system is, $\forall j \in \mathbb{N}$

$$\mathbf{e}_j^m = \mathbf{\Omega}_m \bar{\mathbf{A}}^{j-1}\mathbf{e}_0 + (\mathbf{\Psi}_m + \mathbf{\Omega}_m \bar{\mathbf{B}}^{j-1})(\mathbf{r} - \mathbf{d}) \quad (57)$$

where $\bar{\mathbf{A}}^\ell$ is as in (55) for $\ell \in \mathbb{N}_0$ and $\bar{\mathbf{B}}^\ell$ as in (56) for $\ell \in \mathbb{N}_0$. Moreover, all agents are assumed to start from the same initial trajectory, i.e., $\forall m \in \mathcal{M}$, $\mathbf{e}_0^m = \mathbf{e}_0$.

*Proof:* The proof follows directly from incorporating (54) into (18). ∎

The sequence of best performers, i.e., $(f_0, f_1, \ldots)$, can be calculated for the problem at hand, by recursively solving an optimal control problem. The following proposition formalizes this statement.

*Proposition 5:* The best performer agent at iteration $j \in \mathbb{N}_0$, i.e., $f_j$, can be computed by solving the following optimization problem, $\forall j \in \mathbb{N}_0$:

$$\begin{aligned}f_j = \underset{m\in\mathcal{M}}{\arg\min} \quad & \mathbf{e}_0^T(\mathbf{\Omega}_m\bar{\mathbf{A}}^{j-1})^T(\mathbf{\Omega}_m\bar{\mathbf{A}}^{j-1})\mathbf{e}_0 \\ & + (\mathbf{r}-\mathbf{d})^T(\mathbf{\Psi}_m + \mathbf{\Omega}_m\bar{\mathbf{B}}^{j-1})^T(\mathbf{\Psi}_m + \mathbf{\Omega}_m\bar{\mathbf{B}}^{j-1})(\mathbf{r}-\mathbf{d}) \\ & + 2\mathbf{e}_0^T(\mathbf{\Omega}_m\bar{\mathbf{A}}^{j-1})^T(\mathbf{\Psi}_m + \mathbf{\Omega}_m\bar{\mathbf{B}}^{j-1})(\mathbf{r}-\mathbf{d})\end{aligned} \quad (58)$$

where $\bar{\mathbf{A}}^j$ and $\bar{\mathbf{B}}^j$ are defined as in Proposition 4.

*Proof:* By definition of Euclidean norm

$$\|\mathbf{e}_j^m\|^2 = \mathbf{e}_j^{m^T}\mathbf{e}_j^m.$$

This latter can be expanded by incorporating (57), thus obtaining

$$\begin{aligned}\|\mathbf{e}_j^m\|^2 = & \mathbf{e}_0^T(\mathbf{\Omega}_m\bar{\mathbf{A}}^{j-1})^T(\mathbf{\Omega}_m\bar{\mathbf{A}}^{j-1})\mathbf{e}_0 \\ & + (\mathbf{r}-\mathbf{d})^T(\mathbf{\Psi}_m+\mathbf{\Omega}_m\bar{\mathbf{B}}^{j-1})^T(\mathbf{\Psi}_m+\mathbf{\Omega}_m\bar{\mathbf{B}}^{j-1})(\mathbf{r}-\mathbf{d}) \\ & + 2\mathbf{e}_0^T(\mathbf{\Omega}_m\bar{\mathbf{A}}^{j-1})^T(\mathbf{\Psi}_m+\mathbf{\Omega}_m\bar{\mathbf{B}}^{j-1})(\mathbf{r}-\mathbf{d}).\end{aligned} \quad (59)$$

Consider (14). As any norm yields a nonnegative real number, (14) can be rewritten as

$$\forall j \in \mathbb{N}_0, \quad f_j = \underset{m\in\mathcal{M}}{\arg\min}\|\mathbf{e}_j^m\|^2. \quad (60)$$

By merging (59) into (60), the proof is concluded. ∎

*Remark 7:* Both $\bar{\mathbf{A}}^j$ and $\bar{\mathbf{B}}^j$ depend on the sequence $(f_0, f_1, \ldots)$ that ultimately depends on $\mathcal{M}$, $\mathbf{e}_0$, and $(\mathbf{r} - \mathbf{d})$.

Proposition 5 gives a tool for computing the sequence of best performers for the CILC problem at hand, given the initial conditions and the specific set $\mathcal{M}$.

*Definition 9:* Let, $\forall j \in \mathbb{N}_0$, $\forall m \in \mathcal{M}$, $F_j^m \in \mathbb{R}$ be defined as

$$\begin{aligned}F_j^m := & \mathbf{e}_0^T(\bar{\mathbf{A}}^{j^T}\bar{\mathbf{A}}^j - \tilde{\mathbf{A}}_m^{j^T}\tilde{\mathbf{A}}_m^j)\mathbf{e}_0 \\ & + (\mathbf{r}-\mathbf{d})^T(\bar{\mathbf{B}}^{j^T}\bar{\mathbf{B}}^j - \tilde{\mathbf{B}}_m^{j^T}\tilde{\mathbf{B}}_m^j)(\mathbf{r}-\mathbf{d}) \\ & + 2(\mathbf{r}-\mathbf{d})^T(\bar{\mathbf{B}}^{j^T}\bar{\mathbf{A}}^j - \tilde{\mathbf{B}}_m^{j^T}\tilde{\mathbf{A}}_m^j)\mathbf{e}_0.\end{aligned} \quad (61)$$

By definition of Euclidean norm, by (54) and (53), one can verify that

$$F_j^m = \|\bar{\mathbf{e}}_j\|^2 - \|\tilde{\mathbf{e}}_j^m\|^2. \quad (62)$$

By Remark 7, also $F_j^m$ depends on $\mathcal{M}$, $\mathbf{e}_0$, and $(\mathbf{r} - \mathbf{d})$.

*Theorem 7:* The CILC is well-performing for some $\mathbf{e}_0 \in \mathbb{R}^N$ and $(\mathbf{r} - \mathbf{d}) \in \mathbb{R}^N$ if and only if

$$\forall j \in \mathbb{N}_0 \quad \forall m \in \mathcal{M}, \quad F_j^m \leq 0.$$

*Proof:* Consider condition (52) for which the CILC is well-performing for some $\mathbf{e}_0 \in \mathbb{R}^N$ and $(\mathbf{r}-\mathbf{d}) \in \mathbb{R}^N$. This condition, by (62), corresponds to having $F_j^m \leq 0$. By incorporating this latter into Definition 8, one concludes the proof. ∎

Note that, trivially, in the special case of $\mathbf{u}_0^m = 0$, $\forall m \in \mathcal{M}$, we have $\mathbf{e}_0 = \mathbf{r} - \mathbf{d}$. This way, (61) can be rewritten as

$$\begin{aligned}F_j^m = (\mathbf{r}-\mathbf{d})^T\big(&\bar{\mathbf{A}}^{j^T}\bar{\mathbf{A}}^j - \tilde{\mathbf{A}}_m^{j^T}\tilde{\mathbf{A}}_m^j + \bar{\mathbf{B}}^{j^T}\bar{\mathbf{B}}^j - \tilde{\mathbf{B}}_m^{j^T}\tilde{\mathbf{B}}_m^j \\ & + 2\bar{\mathbf{B}}^{j^T}\bar{\mathbf{A}}^j - 2\tilde{\mathbf{B}}_m^{j^T}\tilde{\mathbf{A}}_m^j\big)(\mathbf{r}-\mathbf{d}).\end{aligned} \quad (63)$$

*Corollary 2:* If, $\forall m \in \mathcal{M}$, $\mathbf{u}_0^m = 0$, the CILC is well-performing for the given $(\mathbf{r} - \mathbf{d}) \in \mathbb{R}^N$ if and only if

$$\forall j \in \mathbb{N}_0 \quad \forall m \in \mathcal{M}, \quad F_j^m \leq 0$$

where $F_j^m$ is defined as in (63).

*Proof:* The proof follows directly from Theorem 7 and (63). ∎

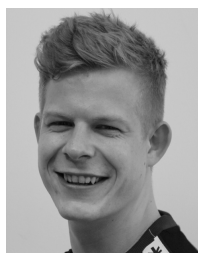

**Michael Meindl** received the M.Sc. degree in mechatronic engineering from Hochschule (HS) Karlsruhe, Karlsruhe, Germany, in 2020. He is currently pursuing the Ph.D. degree with Technische Universität (TU) Berlin, Berlin, Germany.

He is currently a Research Assistant with the Embedded Mechatronics Laboratory, HS Karlsruhe. His research focuses on movement learning in robotic systems with the prime interest of combining methods from the fields of control theory and machine learning.

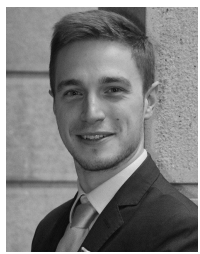

**Fabio Molinari** received the M.Sc. degree in automation and control engineering from the Politecnico di Milano, Milan, Italy, in 2015. He is currently pursuing the Ph.D. degree with Technische Universität Berlin, Berlin, Germany.

He was a Research Assistant with Johannes Kepler Universität Linz, Linz, Austria, until 2016. Since 2016, he has been employed with Technische Universität Berlin. His research focuses on consensus-based control of multi-agent systems over wireless channels (DFG Priority Programme SPP1914).

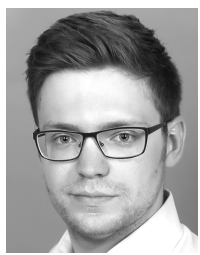

**Dustin Lehmann** received the M.Sc. degree in aeronautics and astronautics from Technische Universität (TU) Berlin, Berlin, Germany, in 2019, where he is currently pursuing the Ph.D. degree with the Science of Intelligence Excellence Cluster.

His research focuses on understanding collective learning in multi-agent systems and applying control theory approaches to collective learning problems.

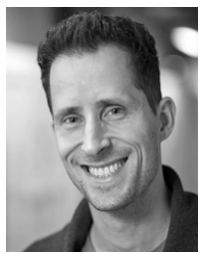

**Thomas Seel** studied engineering cybernetics at Otto von Guericke University (OvGU) Magdeburg, Magdeburg, Germany, and the University of California (UC) at Santa Barbara, Santa Barbara, CA, USA. He received the Ph.D. degree from Technische Universität Berlin, Berlin, Germany, in 2016.

He is currently a Professor with the Department of Artificial Intelligence in Biomedical Engineering, Friedrich-Alexander-Universität Erlangen–Nürnberg, Erlangen, Germany. His research interest includes dynamic inference and learning in biomedical and mechatronic systems.